\documentclass[tightenlines,nofootinbib,superscriptaddress,10pt,titlepage]{revtex4-1}
\usepackage{natbib}
\usepackage{amsmath}
\usepackage{amssymb}
\usepackage{fancyhdr}
\usepackage{url}
\usepackage{hyperref}

\usepackage{siunitx}
\usepackage{suffix}

\makeatletter
\renewcommand{\fnum@table}{\textbf{\tablename~\thetable}}
\renewcommand{\fnum@figure}{\textbf{\figurename~\thefigure}}
\makeatother

\usepackage{suffix}
\usepackage{enumitem}

\newcommand\sectionauthor[1]{\authortoc{#1}\printsectionauthor{#1}}
\WithSuffix\newcommand\sectionauthor*[1]{\printsectionauthor{#1}}

\makeatletter
\newcommand{\printsectionauthor}[1]{%
  {\parindent0pt\vspace*{-10pt}%
  \linespread{1.1}\footnotesize\scshape#1%
  \par\nobreak\vspace*{11pt}}
  \@afterheading%
}
\newcommand{\authortoc}[1]{%
  \addtocontents{toc}{\vspace{-10pt}}%
  \addtocontents{toc}{%
    \protect\contentsline{section}%
    {\hskip1.3em\mdseries\scshape\protect\scriptsize#1}{}{}}
  \addtocontents{toc}{\vskip3pt}%
}
\makeatother

\begin{document}

\title{\Large{Supernova Physics at DUNE} \\ \vspace{5pt}
\normalsize March 11--12$^{th}$, 2016 \\ \vspace{5pt}
\footnotesize Virginia Tech, Blacksburg, VA \\ \vspace{20pt}
}

\author{Artur Ankowski}\affiliation{Center for Neutrino Physics, Department of Physics, Virginia Tech, Blacksburg, VA 24061, USA}
\author{John Beacom}\affiliation{Center for Cosmology and AstroParticle Physics, Ohio State University, Columbus, OH 43210, USA}\affiliation{Department of Physics, Ohio State University, Columbus, OH 43210, USA}\affiliation{Department of Astronomy, Ohio State University, Columbus, OH 43210, USA}
\author{Omar Benhar}\affiliation{INFN and Department of Physics, ``Sapienza'' Universit\'a di Roma, I-00185 Roma, Italy}\affiliation{Center for Neutrino Physics, Department of Physics, Virginia Tech, Blacksburg, VA 24061, USA}
\author{Sun Chen}\affiliation{Center for Neutrino Physics, Department of Physics, Virginia Tech, Blacksburg, VA 24061, USA}
\author{JJ Cherry}\affiliation{Center for Neutrino Physics, Department of Physics, Virginia Tech, Blacksburg, VA 24061, USA}
\author{Yanou Cui}\affiliation{Department of Physics and Astronomy, University of California, Riverside, CA 92521, USA}\affiliation{Perimeter Institute, 31 Caroline Street, North Waterloo, Ontario N2L 2Y5, Canada}\affiliation{Maryland Center for Fundamental Physics, University of Maryland, College Park, MD 20742, USA}
\author{Alexander Friedland}\affiliation{SLAC, Stanford University, Menlo Park, CA 94025 USA}
\author{Ines Gil-Botella}\affiliation{CIEMAT, Department of Basic Research, Av.~Complutense, 40, 28040 Madrid, Spain}
\author{Alireza Haghighat}\affiliation{Nuclear Engineering Program, Mechanical Engineering, Virginia Tech Research Center, Arlington, VA 22203, USA}
\author{Shunsaku Horiuchi}\email{horiuchi@vt.edu}\affiliation{Center for Neutrino Physics, Department of Physics, Virginia Tech, Blacksburg, VA 24061, USA}
\author{Patrick Huber}\affiliation{Center for Neutrino Physics, Department of Physics, Virginia Tech, Blacksburg, VA 24061, USA}
\author{James Kneller}\affiliation{Department of Physics, North Carolina State University, Raleigh, NC 27695, USA}
\author{Ranjan Laha}\affiliation{Kavli Institute for Particle Astrophysics and Cosmology (KIPAC),
Department of Physics, Stanford University, Stanford, CA 94035, USA}
\affiliation{SLAC, Stanford University, Menlo Park, CA 94025 USA}
\author{Shirley Li}\affiliation{Center for Cosmology and AstroParticle Physics, Ohio State University, Columbus, OH 43210, USA}
\author{Jonathan Link}\affiliation{Center for Neutrino Physics, Department of Physics, Virginia Tech, Blacksburg, VA 24061, USA}
\author{Alessandro Lovato}\affiliation{Physics Division, Argonne National Laboratory, Argonne, IL 60439, USA}
\author{Oscar Macias}\affiliation{Center for Neutrino Physics, Department of Physics, Virginia Tech, Blacksburg, VA 24061, USA}
\author{Camillo Mariani}\affiliation{Center for Neutrino Physics, Department of Physics, Virginia Tech, Blacksburg, VA 24061, USA}
\author{Anthony Mezzacappa}\affiliation{Department of Physics and Astronomy, University of Tennessee, Knoxville, TN 37996, USA}
\author{Evan O'Connor}\affiliation{Department of Physics, North Carolina State University, Raleigh, NC 27695, USA}
\author{Erin O'Sullivan}\affiliation{Department of Physics, Duke University, Durham NC 27708, USA}
\author{Andre Rubbia}\affiliation{ETH Zurich, 101 Ramistrasse, CH-8092 Zurich, Switzerland}
\author{Kate Scholberg}\affiliation{Department of Physics, Duke University, Durham NC 27708, USA}
\author{Tatsu Takeuchi}\affiliation{Center for Neutrino Physics, Department of Physics, Virginia Tech, Blacksburg, VA 24061, USA}


\begin{frame}{}

\vspace{400pt}
\noindent
The DUNE/LBNF program aims to address key questions in neutrino physics and astroparticle physics. Realizing DUNE's potential to reconstruct low-energy particles in the 10--100 MeV energy range will bring significant benefits for all DUNE's science goals. In neutrino physics, low-energy sensitivity will improve neutrino energy reconstruction in the GeV range relevant for the kinematics of DUNE's long-baseline oscillation program. In astroparticle physics, low-energy capabilities will make DUNE's far detectors the world's best apparatus for studying the electron-neutrino flux from a supernova. This will open a new window to unrivaled studies of the dynamics and neutronization of a star's central core in real time, the potential discovery of the neutrino mass hierarchy, provide new sensitivity to physics beyond the Standard Model, and evidence of neutrino quantum-coherence effects. The same capabilities will also provide new sensitivity to `boosted dark matter' models that are not observable in traditional direct dark matter detectors.
 \end{frame}

\vspace{-540pt}

\maketitle


\newpage

\section{The DUNE program}\label{sec:dune}

The high-energy physics (HEP) community is currently engaged in implementing the recommendations of the P5~\cite{ParticlePhysicsProjectPrioritizationPanel(P5):2014pwa} report, a central component of which is the Deep Underground Neutrino Experiment (DUNE)/ Long-Baseline Neutrino Facility (LBNF) program.  The DUNE far detector, underground at the Sanford Underground Research Facility (SURF) in South Dakota, will comprise four 10-kton liquid argon (LAr) time projection chambers~(TPCs)~\cite{Acciarri:2016crz,Acciarri:2015uup}. These detectors will be coupled with a $\sim$GeV-neutrino-energy-scale, 1.2-MW beam from Fermilab.  According to current DUNE collaboration strategy, the first two underground detector modules are expected to be commissioned in 2024.  The first 10-kton time-projection-chamber module will make use of single-phase technology, while dual-phase technology is a candidate for subsequent modules~\cite{Acciarri:2016ooe}.

The primary physics aims of DUNE include the following: ($i$) determination of neutrino mass ordering, ($ii$) measurement of CP-violating phase $\delta_{\rm CP}$ and precision tests of the three-flavor oscillation paradigm using long-baseline $\nu_\mu \rightarrow \nu_e$ and $\bar{\nu}_\mu \rightarrow \bar{\nu}_e$ flavor transitions, ($iii$) search for nucleon decay, and ($iv$) observation of the burst of neutrinos from a core-collapse supernova.  Ancillary science topics with the far detector include atmospheric neutrinos, searches for non-standard physics, and dark matter searches.

The last two primary physics goals---proton decay and supernova-neutrino detection---are enabled by the siting of the DUNE detectors at the 4850-ft (4300 m.w.e.) level at SURF.  The overburden means cosmogenic backgrounds are dramatically reduced.  The workshop ``Supernova Physics at DUNE'' focused on topics of HEP and astrophysics related to core-collapse supernova-neutrino signals and the development of DUNE detector requirements to ensure adequate capabilities in the event of a Galactic supernova.

\subsection{Importance of low-energy capabilities for DUNE science}\label{sec:dune:motivation}

A stellar core collapse releases nearly all the binding energy of its compact remnant in the form of a brilliant pulse of neutrinos of all flavors, emitted over a few tens of seconds with energies from a few MeV to a few tens of MeV. The $\nu_e$ component of the signal has promising potential to reveal particularly interesting HEP and astrophysics issues, e.g., the neutrino mass hierarchy, the dynamics of the collapsing core, the mechanism of the supernova explosion, as well as to probe physics beyond the Standard Model (more details are available later, in Section \ref{sec:ccsn}). While existing and near-future large water and scintillator detectors will be sensitive primarily to the electron \textit{anti}neutrino component of the flux, liquid argon is uniquely sensitive to the \textit{electron} neutrino component, primarily via the charged-current interaction, $$\nu_e+{}^{40}{\rm Ar}\rightarrow e^{-}+{}^{40}{\rm K}^{*}.$$ In principle, LAr TPCs have the advantage of fine-grained tracking of charged particles in the detector. Of current and planned experiments worldwide, DUNE is the only detector with an expected high-statistics, clean $\nu_e$ signal. This makes DUNE an important detector in the complementarity of present and future neutrino detectors~\cite{Scholberg:2012id}. It is therefore of great importance to ensure that detector requirements for building DUNE are such that we can extract relevant supernova-neutrino physics.  Core-collapse supernovae within the Milky Way are expected only every 30 years or so; hence it is critical to be well prepared.

It should also be emphasized that DUNE's low-energy capabilities will address a major systematic for DUNE's science program. For the success of the oscillation program of DUNE, an accurate neutrino energy reconstruction is a prerequisite. It is important to note that a single pion below the detection threshold, assumed to be 100 MeV in the Fast Monte Carlo~\cite{Acciarri:2015uup}, may bias the reconstructed energy by up to ~240 MeV, and pion production is the dominant interaction mechanism at the DUNE kinematics. As a consequence, the sensitivity to 4.1-MeV muons produced in pion decay at rest (lifetime 26.0 ns) or to Michel electrons---of energies below 52.8 MeV---from subsequent muon decay at rest (lifetime 2.20 $\mu$s) would greatly improve the accuracy of neutrino-energy reconstruction in the GeV region, supporting the oscillation program as well as improving sensitivity to astrophysical neutrinos in DUNE.

\subsection{Low-energy capabilities of DUNE}\label{sec:dune:lowE}

Beam and atmospheric neutrinos to be used for oscillation physics are mostly in the few-GeV range.  At these energies, it is easy to identify events (especially with the beam for which a beam-timing trigger should be available).  For atmospheric neutrinos there is no trigger, but energetic events look distinctive; the challenge for these events is reconstruction of momenta and energies of all the final-state products (which can be complex, since multiple nucleons may be involved in the scattering) and particle identification in order to deduce the energy and flavor of the interacting neutrino.

In contrast, core-collapse supernova neutrinos are typically in the few to few tens of MeV regime.  At these energies, one still wants to reconstruct the incoming neutrino properties, but the final states are different--- they involve leptons for charged-current events, and possibly nuclear de-excitation products.  The products produce only small stub-like tracks and blips, and hence information from them is dependent on fine-grained resolution.  The signal can be vulnerable to radioactive and cosmogenic background which also produces low-energy events. Like atmospheric neutrinos and nucleon decay events, supernova burst neutrinos do not have a beam trigger; therefore another experimental challenge is capability to trigger on the burst. An advantage, however, for detection of supernova burst events, at least for a nearby supernova, is that they all arrive within a few tens of seconds,  which creates a characteristic signal for triggering and a significant improvement in signal to background. In spite of these advantages, the individual supernova signal events are low enough in energy, and low-energy backgrounds potentially copious enough, that efficient triggering may be challenging.  More work is needed to understand backgrounds and trigger algorithms for a realistic detector configuration.

\begin{table}[b]
\begin{center}
\begin{tabular}{|c|c|c|c|}
\hline
 & & & \\
Detector goal & Value & Main detector systems & Purpose \\
& & & \\ \hline \hline
Trigger efficiency & $>90$\% & Trigger/DAQ \& & SN burst \\
for 5--100 MeV interactions &  & Photon detectors &  \\ \hline
Data acceptance without loss & Non-zero  & DAQ & SN burst \\
\& buffer for at least 2 min & suppression &  &  \\ \hline
Vertex resolution & $\sim$ cm & TPC \& & Background rejection \\
  &  &  Photon detectors  & \\ \hline
Reconstruction of cosmic muons &  & TPC \& & Background rejection \\
and associated radiation &  & Photon detectors  &  \\ \hline
Reconstruction efficiency & $\sim$80\% & TPC \& & Flavor-energy features \\
for 5 MeV events &  & Photon detectors  & in the SN spectrum \\ \hline
Particle identification & & TPC \& & Identif. of $\gamma$ cascades from \\
  &  &  Photon detectors  & low-E $\nu$ int. \& flavor tagging \\ \hline
Energy resolution for & $<10$\% & TPC \& & Features in the \\
5--100 MeV events &  & Photon detectors  & SN $\nu$ spectrum \\ \hline
Absolute time resolution & $< 1$ ms & DAQ \& & SN burst \& \\
 & & Photon detectors & Energy resolution \\ \hline
Angular resolution & $< 20\degree$ & TPC & Event direction \\ \hline
\end{tabular}
\caption{Expected LAr TPC requirements for measuring supernova-neutrino events.}
\label{tab:cap}
\end{center}
\end{table}

Because the neutrino energy spectrum is broad, the better the final-state energies can be measured, the more precisely the neutrino-energy-dependent mixing effect can be measured.  Although the primary final-state lepton will be measured well, it is more challenging to determine the number, nature, and energies of the hadronic particles.  As noted, pions, if they are sub-threshold, can bias the energy reconstruction.  More generally, a combination of techniques may help reconstruct the hadronic state.  Non-relativistic protons may give scintillation signals; non-relativistic neutrons may have inelastic interactions with nuclei; thermal neutrons may have radiative capture signals; and nuclei may be identified through their de-excitations or decays.

In order to measure the time and energy distributions of the different supernova-neutrino flavors, detectors must have good capabilities for neutrino energy resolution, angular and time resolution and particle identification. Table \ref{tab:cap} shows the expected LAr TPC detector capabilities to extract this information and the main detector systems involved. Some components of the DUNE reference design detector systems may need their performance improved to achieve some physics goals. In particular, the photon detectors are critical to many detector goals, from delivering low enough trigger threshold to particle identification and time resolution. Furthermore, in order to quantify detector requirements for DUNE, the theory community is needed to help provide models with physics signatures for simulation studies.

\section{Physics with core-collapse supernova neutrinos}\label{sec:ccsn}

Many terrestrial searches for Standard Model and Beyond the Standard Model physics of neutrinos will be conducted, including with DUNE, but much could also can be learned from studying neutrinos and their properties in supernovae. The simple reason is that in the cores of supernovae the densities, temperatures, magnetic fields, and other conditions can be so high the neutrino is no longer an ephemeral component of the system. As a result, the dynamics of the system, the nucleosynthesis, and the signal we detect with DUNE are very sensitive to the properties of the neutrino. In essence supernovae constitute nature's ultimate neutrino experiment. We first review the status of core-collapse supernova simulation studies, followed by a sample of the many physics a future supernova-neutrino detection can yield, then a summary of the near-future prospects of supernova theory research.

\subsection{Status of core-collapse supernova simulations}\label{sec:ccsn:sim}

Core-collapse supernovae mark the end stage of stellar evolution for stars that begin their lives with the mass of at least $\sim$8--10$\,M_\odot$. Core collapse is triggered when the iron core grows massive enough to surpass the effective Chandrasekhar limit---the point at which the pressure of the electrons, photons, and ions can no longer support the star against gravitational collapse. The collapse ensues and continues until the matter reaches nuclear densities, when the residual strong force between the nucleons is enough to resist gravity and halt collapse. The inertia of the collapsing core causes the newly formed proto-neutron star to overshoot its equilibrium, and the rebounding of the core launches a shock wave that, for a successful core-collapse supernova, will become the supernova shock that unbinds everything but the innermost core of the star. Theoretical work over the past 50+ years in simulating core collapse supernovae has come a long way in elucidating the details of this process. As a broad---but by no means complete---selection of influential work on the core-collapse supernova central engine over the past five years we refer the reader to \citep{2012PTEP.2012aA309J, 2012ApJ...756...84M, 2013RvMP...85..245B,2013ApJ...768..115O, 2013ApJ...770...66H,2014ApJ...785..123C,2014ApJ...786...83T, 2016ApJ...818..123B,Fischer:2009af,Lentz:2015nxa,Kuroda:2015bta,Melson:2015tia,Roberts:2016lzn}. Once the shock is formed, it is driven outward by the intense thermal pressure of the collapsed core, since much of the gravitational binding energy released during the collapse is trapped in internal and thermal energy of the particles. However, at the same time, neutrino emission and the dissociation of heavy nuclei traversing the shock robs the matter of thermal energy, reducing the pressure behind the shock and causing the supernova shock to stall after $\sim$100\,ms. If the shock remains stalled, continued accretion onto the proto-neutron star will eventually cause it to exceed its maximum mass and collapse to a black hole \citep{2011ApJ...730...70O}. However, we know many core collapses must lead to successful core-collapse supernovae \citep{2015PASA...32...16S}. The forefront of supernova theory is understanding the mechanism by which the supernova shock becomes re-energized, propagates away from the core, and unbinds everything but the newly formed neutron star.

The leading theory for shock re-energization is the neutrino mechanism \citep{1966ApJ...143..626C, 1985ApJ...295...14B}. Neutrinos streaming from the core regions interact with and heat material behind the shock, forming the so-called gain region where there is a net increase in the internal energy of the matter due to neutrino interactions. The increase in internal energy leads to an increased thermal pressure in the matter behind the shock, aiding shock expansion. This heating also drives convection that, along with other hydrodynamic instabilities like the standing accretion shock instability (SASI; \citep{2003ApJ...584..971B, 2007ApJ...654.1006F}), are truly multidimensional phenomena and crucial for the success of the neutrino mechanism \citep{2009ApJ...694..664M, 2016ApJ...818..123B}. Many current, state-of-the-art, multidimensional simulations do achieve successful explosions via the neutrino mechanism \citep{2009ApJ...694..664M, 2012ApJ...756...84M, 2014ApJ...786...83T,2015arXiv151107871S,Lentz:2015nxa,Melson:2015tia,2016ApJ...818..123B,Roberts:2016lzn}. Importantly, progress in recent years has demonstrated that a detailed treatment of neutrino transport and neutrino interactions is needed to accurately simulate the core-collapse central engine \citep{2012ApJ...747...73L,2012ApJ...760...94L, 2015ApJ...808L..42M}. This shows that the core-collapse supernova problem itself is highly sensitive to the underlying neutrino physics and consequently, that the detection of neutrinos from the next Galactic supernova will be incredibly informative and transformational for our understanding of how neutrinos are produced, are transported, interact, and behave (i.e., classically and quantum mechanically) under the extreme conditions present in the core-collapse supernova environment.

\subsection{Examples of probes with core-collapse supernova neutrinos}\label{sec:ccsn:eg}

Given the rich neutrino environment in a collapsing core, core-collapse supernovae become ideal laboratories for the exploration of fundamental neutrino physics. Observations of neutrinos from the Sun, the atmosphere, and in terrestrial experiments have shown that neutrinos have mass and therefore can change flavors, from electron to muon and tau, and vice versa. As the neutrino population in a supernova evolves over the course of the explosion, flavor transformations are also expected. Specifically which transformations occur will depend on as-yet-unknown fundamental neutrino physics---e.g., the neutrino mass hierarchy. Consequently, observation of neutrinos from a Galactic core-collapse supernova, in conjunction with detailed predictions of the expected neutrino emissions from sophisticated models, will enable us to pin down these remaining, unknown, fundamental neutrino properties. Observationally, sensitivity to all but $\nu_e$ is available via other detector technologies (Section \ref{sec:broader:flavor}). Hence, DUNE's unique sensitivity to $\nu_e$ is extremely valuable. Below is a selection of some of the HEP and astrophysics topics that will be enabled by DUNE.

\begin{itemize}

\item \textit{Characterizing the supernova}.---It is only by detecting all six flavors of neutrinos---which will have different spectra and time profiles---that a supernova can be completely characterized. This is the only way to robustly measure the supernova's total gravitational binding energy, the process of neutronization of the collapsing core, and the effects of neutrino mixing in extreme conditions. The better the supernova emission is understood, the better tests of neutrino and particle physics will be.

\item \textit{Neutrino mass hierarchy}.---A DUNE detector with the ability to measure the $\nu_e$ flux with good energy resolution has multiple opportunities to determine the neutrino mass hierarchy regardless of the astrophysical uncertainties (i.e., the mass of the progenitor star, neutrino star or black formation, and other inherent uncertainties involved in interpreting the supernova-neutrino signal). Primarily, the ``neutronization burst'' of $\nu_e$ (a close-to standard candle signal during the early phase of the collapse) will only be detectable in the inverted mass hierarchy, oscillations removing the burst in the normal hierarchy \cite{Mirizzi:2015eza}. Secondly, the neutrino oscillation signature of supernovae is the MSW effect induced by the shockwave propagating through the supernovae envelope~\cite{Schirato:2002aa,Tomas:2004aa,2009PhRvL.103g1101G}. The shockwave---ultimately responsible for ejecting the envelope of the star---creates a sharp density gradient which alters the flavor evolution of the $\nu_e$ flux as the explosion progresses. The shock oscillation signatures will be prompt and strong in DUNE if the hierarchy is normal and suppressed / delayed if it is inverted. Finally, spectral swaps resulting from neutrino-neutrino self scattering also encode information about the neutrino mass hierarchy.

\item \textit{Neutrino quantum coherence effect}.---Neutrino emission from supernovae offers a unique opportunity to probe physics at temperatures and densities which are unaccessible by any terrestrial experiment. The DUNE detector itself is indispensable in understanding the oscillation signal of a Galactic supernova due to its direct sensitivity to the $\nu_e$ flux.  One of the prime science targets which DUNE is capable of detecting is the presence of spectral \lq\lq swaps\rq\rq\ in the $\nu_e$ signal~\cite{Duan06a}. Such swaps are direct evidence of neutrino-neutrino self scattering~\cite{Pantaleone:1992xh,Pantaleone:1992eq,Pantaleone:1995aa}, a macroscopic quantum coherence phenomenon. Due to the preponderance of $\nu_e$ flux in the supernova-neutrino emission, swaps are expected to be observable by DUNE for the bulk of the supernova-neutrino burst.

\item \textit{Beyond Standard Model physics}.---The pursuit of Beyond the Standard Model (BSM) physics is a major goal of current nuclear and HEP research. BSM scenarios with supernovae neutrinos have been considered over the years and shown to alter expectations relative to the Standard Model predictions. For example, the effect of sterile neutrinos in supernovae has been considered in \cite{1999PhRvC..59.2873M,2006PhRvD..73i3007B,2012JCAP...01..013T,2014PhRvD..90j3007W,2014PhRvD..89f1303W,Arguelles:2016uwb}; neutrino magnetic moments were studied by \cite{1988PhRvL..61...27B,1999PhLB..470..157M,2007JCAP...09..016B,2012JCAP...10..027D}; and Non-Standard Interactions (NSI) of neutrinos with matter have been studied by \cite{PhysRevD.76.053001,2008PhRvD..78k3004B,2010PhRvD..81f3003E,arXiv1605.04903}. Arguments based on novel energy-loss processes have led to competitive limits on light particles such as axions \cite{Raffelt:1987yt}, which stand to be improved by a future full six flavor detection. The prospects of DUNE uncovering BSM physics in a supernova or compact object merger neutrino signal is a very exciting possibility that need to be exploited as fully as possible.

\item \textit{Multi-messenger astrophysics}.---The temporal, spectral, and angular information of supernova $\nu_e$ flux will be a key part of future multi-messenger astronomy that includes gravitational waves and the electromagnetic spectrum from radio to gamma rays \cite{Nakamura:2016kkl}. These will collectively help reveal the stellar structure, core-collapse dynamics, and formation of neutron stars or black holes. As a specific example focusing on the $\nu_e$ signal, the timing and energy dependence of the shock effect on the $\nu_e$ reveal information about the interior structure of the exploding star which cannot be measured in any other way. Once the shock flavor oscillation signature has passed, the turbulent convection of matter behind the shock will also produce a unique $\nu_e$ flavor oscillation signature which DUNE will be sensitive to~\cite{Friedland:2006lr,2010PhRvD..82l3004K,2013PhRvD..88b3008L}.  Turbulence will leave imprints in the signal that allow DUNE to directly measure the presence of the SASI and shed light on the supernovae explosion mechanism. It should be emphasized that for DUNE to accurately reconstruct the heavy lepton flavor neutrino spectrum, it will be necessary to accurately tag the de-excitation gamma rays from neutral-current $\nu$-Ar interactions. 

\end{itemize}

\subsection{The next decade of supernova theory}\label{sec:ccsn:future}

As simulations of the core-collapse supernova central engine progress forward, we can expect complete neutrino signal predictions covering the entire time-span of the supernova. This range will cover the pre-collapse stages up to the neutronization burst, the accretion phase---including self-consistent explosions---and the proto-neutron star cooling phase. The neutrinos from each of these phases relay detailed information about the underlying stellar core hydrodynamics and thermodynamics. These neutrinos are one of the only ways, and arguably the best way, to directly observe the inner workings of a core-collapse supernova. However, to truly learn about the central engine, these predictions need to be tested, and the only way to truly test them will be in the context of an actual Galactic core-collapse supernova observation (the conditions in a core-collapse supernova cannot be reproduced in terrestrial laboratories). By combining the detection of neutrinos from the next Galactic supernova with the predictions of detailed models from the several supernova modeling groups worldwide, only then will we be able to unravel the complexities of the core-collapse supernova central engine and the remaining associated fundamental neutrino physics not yet in hand. Consequently, ensuring that we can detect this signal with as much detail as possible is of the utmost importance to both astrophysics and neutrino physics, and by extension the Standard Model of elementary particles.

\section{Feasibility}\label{sec:feasibility}

The technical LAr TPC detector capabilities required for measuring a core-collapse supernova-neutrino burst were summarized in Table \ref{tab:cap}. Below we discuss two additional important considerations that must be understood and controlled: the nuclear physics of neutrino-argon interactions, and detector backgrounds.

\subsection{Cross sections}\label{sec:feasibility:crosssection}

More work is necessary on both the experimental and theoretical fronts before we can determine a fully-informed set of detector requirements for DUNE.

On the theoretical side, there are two areas for which additional input would be helpful.  First, more detailed signals and simulations are highly desirable, as described in the sections above.  Second, and very important, theoretical input is needed to achieve a better understanding of neutrino interactions in the supernova energy regime. While a good deal of attention is (rightly) being paid to neutrino interactions in the $\sim$ GeV range, relevant to beam and atmospheric oscillation studies, there is relatively little theoretical literature addressing the few to few tens of MeV region, where the impulse approximation is expected to break down and different effects of nuclear dynamics become important.  To evaluate the observability of signals, experimentalists not only need to know interaction rates, but they must also have reliable models of the interaction products on argon. In fact, understanding of a supernova burst signal in detectors worldwide will require that these models be also extended to other nuclei besides argon.

The dominant reactions involving supernova $\nu_e$ and $\bar{\nu}_e$ in liquid argon detectors are the charged-current absorption processes $\nu_e + {{^{40}}{\rm Ar}} \to e^- + {{^{40}}{\rm K}^*}$ and $\bar{\nu}_e + {{^{40}}{\rm Ar}} \to e^+ + {{^{40}}{\rm Cl}^*} $ followed by de-excitation of $^{40}{\rm K}^*$  and $^{40}{\rm Cl}^*$ via $\gamma$-ray emission. The existing theoretical calculations of the corresponding cross sections, some of them exploited to estimate the expected event rates~\cite{Bueno:2003}, have been carried out within the shell model (SM) scheme~\cite{Ormand:1995}, the random phase approximation (RPA)~\cite{Kolbe:2003} and the local density approximation (LDA)~\cite{Sajjad:2004}.

Over the past decade, a substantial amount of  work has been devoted to the application of  theoretical approaches based on realistic models of nuclear dynamics\textemdash such as the Green's Function Monte Carlo (GFMC) computational technique~\cite{GFMC} and Correlated Basis Function (CBF) perturbation theory~\cite{CBF}\textemdash for the study of neutrino-nucleus interactions.

Advanced many-body calculations of the responses of nuclear matter and medium-heavy nuclei to electroweak interactions have been carried out in a broad kinematical range, extending from a few MeV~\cite{Lovato1,Lovato2} to a few GeV~\cite{Ankowski,Rocco,Lovato3}. The emerging pattern clearly indicates that a consistent treatment of the complexity of dynamical effects not taken into account within the independent particle model,  most notably short- and long-range correlations, is called for.

The extension of the existing many-body approaches to the treatment of a nucleus as complex as argon will require a significant effort and a synergy between different schemes, as well as  an efficient use of the leadership-class computer capabilities available at present time. For example, the nuclear matter responses can be exploited to obtain the nuclear cross sections within LDA, and benchmark calculations performed using the computationally-intensive GFMC technique can provide valuable information on the validity of more-approximate computational schemes. While not involving serious conceptual difficulties, however, the achievement of  these goals will require a description of the weak response of nuclear systems with large neutron excess. In this context, a significant role will be played by  the development of the Self Consistent Green's Function  (SCGF) method~\cite{SCGF}, which appears to be well suited to study the properties of neutron-rich nuclei.

Experimental work in this area is also highly desirable.  To date, only one nucleus has had neutrino-interaction rates and products measured in the tens-of-MeV range at the $\sim$10\% level; this is $^{12}$C,  for which charged-current and neutral-current interactions were measured by LSND~\cite{Auerbach:2001hz} and KARMEN~\cite{Armbruster:1998gk}, respectively. There have been no direct measurements of neutrinos on argon, although one can infer neutrino interactions rates using measurements of other processes, such as beta-decay to a final-state mirror nucleus~\cite{Bhattacharya:1998hc} or charge-exchange scattering reactions~\cite{Bhattacharya:2009zz}; however there is currently no consistent framework or fully reliable calculation using these methods~\cite{Karakoc:2014awa}. A promising type of source for direct neutrino measurements is a pion decay-at-rest source, such as the Spallation Neutron Source (SNS)~\cite{Bolozdynya:2012xv}, which produces neutrinos with energies up to 50 MeV with a well-understood spectrum.  Currently at the SNS there are measurements ongoing of coherent elastic neutrino nucleus scattering (also relevent for supernova neutrinos) and neutrino interactions on lead~\cite{Akimov:2015nza}. While there are at the time of this writing no concrete plans, a future program for argon or other nuclei at the SNS or similar sources would be valuable, and perhaps eventually critical for the interpretation of a supernova burst signal.

\subsection{Detector backgrounds}\label{sec:feasibility:bkg}

Although DUNE's depth---4850-ft level at SURF---will significantly reduce cosmic backgrounds, the consideration of backgrounds is essential for understanding the physics feasibility of the DUNE detector. However, they are not well studied. Potential backgrounds include neutron capture processes, radioactive backgrounds, cosmogenics, and electronic noise~\cite{Gehman2014, Botella2016}.

Cosmogenic isotopes produced by cosmic-ray muons are likely to dominate backgrounds above $\sim$5~MeV.  So far, there is only one simulation study on their yields in argon detectors~\cite{Barker:2012nb}.  Detailed studies both in simulations and in measurements are needed to understand the average neutron- and isotope-production rates in LAr.  For detecting supernova-burst neutrinos in particular, the variations in cosmogenic production among different primary muons are crucial.  Such simulation studies are underway~\cite{Li2016},  building on work for Super-Kamiokande \cite{Li:2014sea,Li:2015kpa,Li:2015lxa}. Sufficiently controlled backgrounds can also open DUNE to measurements of steady-state signals which will provide a constant stream of results, e.g., solar neutrinos and the diffuse flux of all past supernova neutrinos.

\section{Broader impact of low-energy capabilities}\label{sec:broader}

\subsection{Multi-flavor neutrino physics}\label{sec:broader:flavor}

Water-Cherenkov detectors have particular historical importance for supernova-neutrino detection. In 1987, the Kamiokande detector observed a handful of neutrino events from a supernova event, resulting in a Nobel Prize in 2002. Water-Cherenkov detectors have the advantage of an inexpensive target material, which means they can typically be much larger in size than those employing other neutrino-detection technologies.

A proposed water-Cherenkov detector that plans to be online around the same time as the DUNE detector is Hyper-Kamiokande (Hyper-K). Hyper-K would be a 516 (374) kton total (fiducial) volume detector located in Japan, with an inner detector of 440 kton usable for detecting a supernova neutrino burst. The detection would provide complementary information to that of DUNE: Hyper-K will measure mainly electron antineutrinos, while DUNE will primarily see electron neutrinos. This means that features in the early times of explosion where $\nu_e$ are emitted, such as the infall and neutronization peak, will be best measured by DUNE. In the late-time of a supernova explosion, all neutrino flavors are emitted and therefore the higher target mass of Hyper-K will result in a better measurement in this phase. Together with JUNO, a 20 kton liquid scintillator detector currently under construction, there will be three large-volume detectors with differing technologies providing highly complementary data.

The increased information on the $\nu_e$, $\bar\nu_e$, and $\nu_x$ fluxes---which can be gleaned by combining the signal from DUNE, Hyper-K, JUNO, and RENO-50---opens a world of possibilities for studying the phenomenon of the supernova explosion.  By comparing the $\nu_e$ and $\bar\nu_e$ fluxes a measurement of the net lepton number of the collapsing core can be made.  Observing the time evolution of the lepton fluence illuminates the neutrino-diffusion timescale, which constrains the equation of state of neutron-star matter.  Confirmation of the \lq\lq neutrino lighthouse\rq\rq\ effect~\cite{Tamborra:2014aa}, where a hot spot on the surface of a rotating proto-neutron star creates an oscillating neutrino signal, will require a temporal comparison of all $\nu_e$, $\bar\nu_e$, and $\nu_x$ fluxes. Beyond the Standard Model physics can also be tested through the potential $\nu$-$\bar\nu$ oscillation signature which would indicate the Majorana nature of neutrinos.  This effect can be created by either neutrino magnetic-moment interactions~\cite{de-Gouvea:2013aa} or quantum kinetic effects~\cite{Cirigliano:2015aa}, but requires the comparison of $\nu_e$ and $\bar\nu_e$ fluxes to establish that $\langle E_{\nu_e} \rangle > \langle E_{\bar\nu_e}\rangle$ with a high degree of statistical significance, thus confirming $\nu$-$\bar\nu$ oscillation.

The addition of complementary detectors also aids in other ways. Since the supernova burst duration is $\sim$10 sec, it can be easily missed due to detector downtime. Also, different detectors will help reduce the systematic uncertainties that will be present in various neutrino-nucleus interactions. The Super-K detector enriched with gadolinium can detect supernova $\nu_e$ via $\nu_e$ + $e^-$ elastic scattering.  If the neutrino temperature is high, then $\nu_{e} + ^{16}$O interaction can also detected in substantial numbers.  The idea behind this is extremely simple: enriched with gadolinium, Super-K can individually distinguish supernova $\bar{\nu}_e$ interactions via inverse beta decay \cite{Beacom:2003nk}.  Once these interactions are individually removed, the $\nu_e$ + $e^{-}$ can be easily isolated via angular cut.  Similarly, this enormous reduction of the background enabled by the addition of gadolinium can also help in detection of $\nu_{e} + ^{16}$O interaction. Super-K can reconstruct the properties of $\nu_e$ spectrum to 20\% using this technique \cite{Laha:2013hva}.  Due to the bigger volume, Hyper-K can improve the limits by a further factor of $\sim$5.  JUNO can reconstruct the supernova $\nu_e$ properties with comparable precision\,\cite{Laha:2014yua}, and neutrino-proton elastic scattering can provide a good handle on $\nu_x$ \cite{Beacom:2002hs,Dasgupta:2011wg}.

\subsection{Astroparticle physics and cosmology}\label{sec:broader:dm}

DUNE also holds great potential to unravel other prominent puzzles in particle physics and cosmology. Among other possibilities, this includes testing the nature of dark matter (DM) via indirect detection of WIMP annihilation into neutrinos, as well as detection of low-mass dark photon/DM that may be produced from $pp$ beam collision \cite{Batell:2009di}. Considering the complexity of our visible Standard-Model sector, however, a non-minimal DM sector is a fairly generic possibility that has drawn rising interest among theorists. Recently, it was pointed out that a small fraction of the cosmic background of DM may be (semi-)relativistic, branded as ``boosted dark matter'' \cite{Agashe:2014yua, Berger:2014sqa}. DUNE will be an ideal detector for such DM searches.

Boosted DM can be created via a number of possible processes, including annihilation and decays, e.g., in a two-component DM model where a heavier DM annihilates into a lighter subleading DM component \cite{Agashe:2014yua,Belanger:2011ww,SungCheon:2008ts,Berger:2014sqa}. When the boosted DM flux arrives at a terrestrial experiment detector, it can scatter off $e^-$ and/or nucleons through a neutral-current type of interaction mediated by a dark photon or $Z'$ boson. As the incoming boosted DM itself is (semi-)relativistic, it tends to give the target $e^-,~p$ in the detector material a hard kick, and thus the outgoing $e^-,~p$ would be typically energetic. In general, regular DM direct detection experiments are not suitable for boosted DM detection due to their small detector size and the focused sensitivity to $e^-$ or nucleon with very low recoil energy of $O(\rm keV)$ (unless DM has very low mass, see \cite{Cherry:2015oca}). In contrast, neutrino experiments can have much larger detector volume and are automatically sensitive to energetic $e^-,~p$ by design, thus are ideal facilities for direct detection of boosted DM.

The phenomenology of boosted DM search has so-far focused on Super-K and IceCube \cite{Agashe:2014yua, Berger:2014sqa}, showing them to have good sensitivity to some parameter space of the boosted DM models in question. However, the Cherenkov threshold cut is especially harsh for elastic proton scattering signals  ($>1.07$ GeV$/c$ $p$ recoil momentum in water) as there is only a small fraction of energy region above the Cherenkov threshold where the nucleon form-factor suppression is not significant \cite{Beacom:2003zu}. On the other hand, if one considers the total scattering cross section without the Cherenkov threshold cut, the proton channel can generally be the leading discovery mode compared to the electron channel. This is particularly true in a large class of models where the dark mediator is leptophobic \cite{Carone:1994aa, Pospelov:2011ha, Tulin:2014tya}, i.e. it only directly interacts with baryons in the SM. Such severe penalty on signal rates is absent in the new generation of LAr detectors such as the ones employed by DUNE, leading to orders of magnitude improvement in signal sensitivity in the proton channel \cite{Agashe:2014yua, Berger:2014sqa}. The ultimate capacity at DUNE will depend on how good the sensitivity to low-energy $e^-,~p$ events end up being. From this point of view, $\sim$10 MeV is a desirable target for the sensitivity to the kinetic energy of protons. Recently, a working group including several experimentalists from the MicroBooNE and DUNE Collaborations and the authors of \cite{Agashe:2014yua, Berger:2014sqa} has been formed. The goal of the working group is to investigate the sensitivity of the new LAr neutrino detectors at DUNE to the boosted DM signals \cite{collab_dm}.

\section{Conclusions}\label{sec:conclusion}

DUNE's neutrino physics and astroparticle physics aims will be enhanced, or in some cases only possible, by attaining DUNE's low energy [${\mathcal O}$(10) MeV] capabilities. The success of DUNE's oscillation program relies heavily on the capability to reconstruct neutrino energy accurately. Sensitivity to low-energy particles---e.g., muon from pion decay or Michel electron from subsequent muon decay---will reduce the amount of missing energy and significantly improve neutrino energy reconstruction in the GeV range, relevant for the kinematics of DUNE. In addition, low-energy capabilities will open a new window to unrivaled studies of neutrino physics and astrophysics with supernova neutrinos. Thanks to its unique sensitivity to electron neutrinos, DUNE's large-volume deep-underground far detectors have the potential to provide invaluable information. In addition to allowing a real-time observation of the dynamics and neutronization of a star's central core to a neutron star (and perhaps even to a black-hole), measurements of the $\nu_e$ signal can lead to the discovery of the neutrino mass hierarchy, provide new sensitivity to physics beyond the Standard Model such as boosted DM, and evidence of neutrino quantum-coherence effects. Therefore, there are wide and far-reaching motivations to ensure the low-energy sensitivity in the design of the DUNE's far detectors.

\newpage

\appendix


\section*{References}

\bibliographystyle{apsrev}
\bibliography{ref_dune}
\end{document}